\documentclass[twocolumn,showpacs,amsmath,nofootinbib,amssymb]{revtex4}

\usepackage{epsfig}
\usepackage{bm}

\newcommand{\p}{\ensuremath{{\rm p}}}
\newcommand{\n}{\ensuremath{{\rm n}}}
\newcommand{\D}{\ensuremath{{\rm D}}}
\newcommand{\T}{\ensuremath{{\rm T}}}
\newcommand{\He}[1][4]{\ensuremath{^{#1}{\rm He}}}
\newcommand{\Li}[1][7]{\ensuremath{^{#1}{\rm Li}}}
\newcommand{\Be}[1][7]{\ensuremath{^{#1}{\rm Be}}}

\begin{document}

\title{BBN and $\Lambda_{QCD}$}

\author{James P. Kneller}
\email{Jim_Kneller@ncsu.edu}
\affiliation{Department of Physics, North Carolina State University, Raleigh, North Carolina 27695-8202}

\author{Gail C. McLaughlin}
\email{Gail_McLaughlin@ncsu.edu}
\affiliation{Department of Physics, North Carolina State University, Raleigh, North Carolina 27695-8202}

\date{\today}

\begin{abstract}
\end{abstract}


\begin{abstract}

Big Bang Nucleosynthesis (BBN) has increasingly become the tool of choice for investigating the permitted variation of
fundamental constants during the earliest epochs of the Universe. Here we present a BBN calculation that has been
modified to permit changes in the QCD scale, $\Lambda_{QCD}$. The primary effects of changing the QCD scale upon BBN
are through the deuteron binding energy, $B_{\D}$, and the neutron-proton mass difference, $\delta m_{np}$, which both
play crucial roles in determining the primordial abundances. In this paper, we show how a simplified BBN calculation
allows us to restrict the nuclear data we need to just $B_{\D}$ and $\delta m_{np}$ yet still gives useful results so
that any variation in $\Lambda_{QCD}$ may be constrained via the corresponding shifts in $B_{\D}$ and $\delta m_{np}$
by using the current estimates of the primordial deuterium abundance and helium mass fraction. The simplification
predicts the helium-4 and deuterium abundances to within 1\% and 50\% respectively when compared with the results of a
standard BBN code. But $\Lambda_{QCD}$ also affects much of remaining required nuclear input so this method introduces
a systematic error into the calculation and we find a degeneracy between $B_{\D}$ and $\delta m_{np}$. We show how
increased understanding of the relationship of the pion mass and/or $B_{\D}$ to other nuclear parameters, such as the
binding energy of tritium and the cross section of $\T+\D\rightarrow \He+n$, would yield constraints upon any change in
$B_{\D}$ and $\delta m_{np}$ at the 10\% level.

\end{abstract}

\maketitle

\newpage

\section{Introduction}
\label{sec:intro}

Speculation that fundamental constants may vary with time began as early as the 1930's \cite{dirac} and, although there
is no mechanism for such time variation in the context of the standard model of particle physics, recent observations
have created renewed interest in this idea. Webb et al. \cite{Webb:1998cq} report observations of quasar absorption
lines at redshift of z=1-2 that suggest the fine structure constant, $\alpha$, may have been smaller at this time. The
effect is at the level of one part in $10^5$ and further analysis of a new data set by the same group gives similar
results \cite{Webb:2000mn} although Bahcall, Steinhardt and Schlegel \cite{Bahcall:2003rh} using a different analysis
method find a different limit.

Independently of these observations however, it is interesting to consider what happens if the fundamental constants
were different at earlier time than they are today. There are many suggestions for beyond the standard model theories
that could accommodate time variation in the fundamental constants and link the changes in some constants to others,
e.g. Langacker \cite{Langacker:2001td}, though in general studies have typically derived constraints on one constant
when all the others are fixed. If these constants were different at an early epoch than they are today, their relative
shifts can only be determined by the underlying theory which causes the changes.  In the context of grand unifies
theories, all couplings are correlated via the GUT scale parameter, which varies with time, and relative ratios are
determined by the renormalization group equations.
Langacker et al. \cite{Langacker:2001td} (see also \cite{Calmet:2002ja,Calmet:2001nu}, find that
\begin{equation}
{\Delta \Lambda_{QCD} \over \Lambda_{QCD}} \sim 30 {\Delta \alpha \over \alpha}. \label{eq:30deltaalpha}
\end{equation}
The relationship between the shift in the fine structure constant and the vacuum expectation value of the Higgs boson
is very model dependent even in the context of grand unified theories. For some supersymmetric theories it is as large
as ${\Delta v / v} \sim 70 {\Delta \alpha / \alpha}$. While the limits on the variation of the fine structure constant
derived from quasar spectra (which were formed $\sim$ 1 billion years after the Big Bang) are severe and would seem to
limit the amount we could expect in a variation of $\Lambda_{QCD}$ at that epoch (for example from equation
\ref{eq:30deltaalpha}) they by no means rule out the possibility of much larger changes in $\Lambda_{QCD}$, $\alpha$ or
indeed any other constant at the much earlier phase of BBN unless a model is provided that allows one to calculate how
their value at one epoch is related to the value at another.

To constrain the permitted variations one can look to several places to confront theory with data and the strongest
constraints upon the time variation of fundamental constants are expected to emerge when the observables come from
events in the distant past. Such places include: the Oklo nuclear reactor, quasar absorption spectra, the Cosmic
Microwave Background and Big Bang nucleosynthesis. There have been a number of studies that consider the variation of
particular fundamental constants in these scenarios. Data from The Cosmic Microwave Background anisotropy measurements
and BBN were used by Yoo and Scherrer \cite{Yoo:2002vw} for the Higgs vacuum expectation value while the effect of
changing the fine structure on Big Bang Nucleosynthesis has been extensively studied by Bergstr\"{o}m, Iguri and
Rubinstein \cite{Bergstrom:1999wm} and Nollett and Lopez \cite{Nollett:2002da}. Flambaum and Shuryak
\cite{Flambaum:2002de,Flambaum:2002wq} give constraints upon the quark masses and $\Lambda_{QCD}$ after using BBN and
limits derived from the Oklo nuclear reactor have also been studied \cite{Fujii:2002hc,Flambaum:2002wq}. In a study
that permitted changes in all the gauge and Yukawa couplings by relating them to the evolution of a single scalar field
(the `dilaton') Ichikawa and Kawasaki \cite{Ichikawa:2002bt} again use BBN to limit the variations of these couplings.  Fundamental coupling constraints
on BBN were also studied by \cite{Campbell:1994bf}.

In this paper we revisit the study of the influence of the strong coupling on Big Bang Nucleosynthesis yields. BBN
occurs in the first few minutes and hours after the Big Bang and hence we would naively expect the maximal difference
from the current values of the fundamental constants if indeed they vary with time. A variation in $\Lambda_{QCD}$ is
much more difficult to implement than changes in most other constants because of the uncertainty of how the difference
would impact the nuclear data required to make a prediction. Nevertheless, that is our intention in this paper. In the
next section we show that of all the inputs to the calculation, it is the deuteron binding energy, $B_{\D}$, and the
neutron-proton mass difference, $\delta m_{np}$ that play the most crucial role in determining the primordial helium
and deuterium abundances - the two nuclei with, currently, the two most accurately known primordial abundances with
which to compare the calculation. In the following section we discuss how the variation in fundamental parameters is
related to the deuteron binding energy and the neutron-proton mass difference and discuss how these quantities enter
into the important reaction rates. In the fourth section we describe a calculation of Big Bang nucleosynthesis with a
modified reaction network and show explicitly the effects of $\delta m_{np}$ and $B_{\D}$ on abundance yields before
deriving constraints on these quantities. In the last section we give our conclusions and point to where further
understanding of the interactions/structure of nuclei can improve our results.


\section{Standard BBN}
\label{sec:stanbbn}

BBN represents the marriage of nuclear physics with cosmology and, in comparison to the majority of nucleosynthesis
settings, the calculation is relatively simple since there are no production zones to deal with (the whole Universe
participates) and no spatial gradients of any kind that lead to a transport of entropy, momentum or mass. But the
continual dilution and cooling of the cosmic fluid certainly does not mean the Universe is in a steady-state during
these early phases. The expansion is driven by the energy density of the relativistic particles, the nucleons/nuclei
represent only a small fraction of the total, and the inexorable decrease in temperature and density means that the
nuclear reactions can occur for only a brief period. Despite their cosmological inconsequence the (inferred) abundances
of the nuclei at the end of BBN represent the best set of observables that probe sub-horizon scale physics at this
early epoch. Like all nucleosynthesis mechanisms BBN is sensitive to three key characteristics of the setting: the
duration, the energy available and the interactions between the nuclei. Simplistically the primordial abundances are
determined in three (somewhat) distinct phases whose boundaries are determined by just two nuclear parameters: the
neutron-proton mass difference $\delta m_{np}$ and the binding energy of the deuteron $B_{\D}$. In each, the behavior
of the nuclei is distinguished by their interactions and, in order to set the stage for later discussions, we shall
skip briefly through each pointing out the importance of these two parameters.


\subsection{$\delta m_{np}$: Weak Equilibrium}

Throughout the entire evolution of the baryons in BBN it is the relativistic particles that drive the expansion,
provide the thermal bath in which the nuclear reactions take place, and set the time-temperature relationship. At a
sufficiently early epoch all particles are in thermal contact through electromagnetic and/or weak interactions and so
possess a common temperature $T$. At a temperature of $10\;{\rm MeV}$, a common initial temperature for BBN
calculations, the radiation includes: photons, electron/positrons and three light neutrinos. The inconsequence of the
nucleons/nuclei for cosmology at this epoch is implied by the smallness of the ratio of baryon and photon number
densities, denoted by $\eta$, which is $\eta = n_{B}/n_{\gamma} \sim {\cal O}(10^{-10})$. Any chemical potential of the
photons is driven to zero from such rapid processes as double Compton scattering, in the absence of any (significant,
non-thermal) source of neutrinos the ratio $\xi_{i} = \mu_{i}/T_{i}$ is constant, and the chemical potential of the
electrons/positrons is set by the proton density and is therefore extremely small, $\xi_{e} \sim \eta$ \cite{kolb}. The
energy density of the Universe is thus determined from the common temperature and the chemical potential of each
relativistic fluid component so that the expansion rate of the Universe, denoted by the Hubble parameter $H$, is simply
\begin{equation}
H^{2} = \frac{8\pi G_{N}}{3} \, \sum_{i} \rho_{i}(T_{i},\mu_{i})
\end{equation}
and the age of the Universe is related to the temperature through
\begin{equation}
t\,T^{2} = 0.74\;{\rm MeV^{2}\,s}, \label{tpre}.
\end{equation}
when all neutrino chemical potentials are zero.

The neutrons and protons are also held in chemical equilibrium via such weak interactions as
\begin{equation}
\n + \bar{e} \leftrightarrow \p + \bar{\nu}_{e}.
\end{equation}
From the requirement $\mu_{\n} - \mu_{e} = \mu_{\p} - \mu_{\nu_{e}}$ the neutron to proton ratio $F$ is
\begin{equation}
F = \frac{n_{\n}}{n_{\p}} =  \exp \left[ -\frac{\delta m_{np}}{T} + \xi_{e} - \xi_{\nu_{e}} \right].
\end{equation}

As the Universe cools the plethora of equilibria established above $10\;{\rm MeV}$ is broken by the increasingly
infrequent weak interactions. The simplicity of a single temperature for every fluid constituent falters when the
neutrinos `decouple'. The ebb of the weak interactions occurs when the interaction timescale becomes longer than the
age of the Universe so that the neutrino decoupling temperature represents the point at which the two are equal. Below
this temperature a neutrino is, on average, unlikely to ever experience another interaction that would allow energy to
be transferred from one fluid component to another. For $\nu_{\mu}$ and $\nu_{\tau}$ the decoupling temperature is $T
\sim 3.3\;{\rm MeV}$, for $\nu_{e}$ it is slightly lower at $T \sim 2\;{\rm MeV}$ because $\nu_{e}$ may also interact
with the electron/positron fluid via $W^{\pm}$ exchange whereas the other two flavors do not. Despite the fact that the
three neutrinos no longer interact with any other fluid component there is no change in the evolution of the Universe.
Since there was no change in the number of degrees of freedom in any of the fluid components during decoupling the
three flavors all possess the same temperature $T_{\nu}$ equal to the electromagnetic temperature $T_{\gamma}$.
Neutrino decoupling has no direct impact upon the nucleons at this point in their evolution, it's importance is
resurrected during a later epoch.

Concurrent with neutrino decoupling the nucleons also undergo an equilibrium crisis. The neutron/proton interconversion
reaction $\n \leftrightarrow \p$,  which actually represents three processes,
\begin{subequations}
\begin{eqnarray}
\n & \leftrightarrow & \p + e + \bar{\nu}_{e} \label{eq:ndecay}\\
\n+\bar{e} & \leftrightarrow & \p + \bar{\nu}_{e} \label{eq:nebar}\\
\n+\nu_{e} & \leftrightarrow & \p + e, \label{eq:nnue}
\end{eqnarray}
\end{subequations}
is also governed by the weak interaction and is the only process that can significantly alter the neutron/proton ratio.

At temperatures of $\sim 10\;{\rm MeV}$ the three body reaction $\p + e + \bar{\nu}_{e} \rightarrow \n$ and its
inverse, neutron decay, both occur at a rate smaller than the Hubble parameter and are therefore incapable of affecting
the neutron to proton ratio. In contrast the two-body reactions (\ref{eq:nebar}) and (\ref{eq:nnue}) are sufficiently
rapid initially to establish an equilibrium but below $T_{\gamma}\sim 1.2\;{\rm MeV}$, a temperature we shall denote by
$T_{np}$, the rates become smaller than the Hubble rate and are therefore incapable of maintaining this equilibrium.
The neutron to proton ratio is said to `freeze out' though, in truth, even if all complex nuclei were prohibited and
neutrons did not decay the ratio would not become a constant until much later \cite{BBF1989,S1996}.

The departure from neutron/proton weak equilibrium sets the first milestone in the path to BBN. The boundary between
the equilibrium phases prior to $\sim 1\;{\rm MeV}$ and the non-equilibrium phase that follows is determined
essentially by the single parameter $\delta m_{np}$ and this quantity also determines the freeze-out ratio of neutrons
and protons. The freeze-out ratio sets the upper limit to the number of neutrons that can participate in BBN and will
therefore limit the primordial abundance of every complex nucleus. As the Universe continues to evolve below $\sim
1\;{\rm MeV}$ the $n/p$ ratio decreases from the freeze-out ratio because of both the fading residual two-body
interactions and the emerging importance of neutron decay: the extent to which this limits the number of neutrons that
will go on to participate in BBN proper is determined by the second parameter $B_{\D}$.


\subsection{$B_{\D}$: NSE}

There are several key events that occur during this second stage of the maturing Universe. The first is to the
background fluid when the number/energy density of $e,\bar{e}$ begins to depart from the relativistic value. This
occurs at roughly $T_{\gamma}\sim 800\;{\rm keV}$ and does not cease until $T_{\gamma} \sim 10\;{\rm keV}$ when the
annihilation rate becomes smaller than the Hubble rate. The electron/positron annihilation deposits energy and entropy
only into the photon gas, the neutrinos have decoupled and cannot share in this energy release. The change in the
number of electromagnetic degrees of freedom leads to an increase of $T_{\gamma}$ relative to $T_{\nu}$ in order to
maintain the entropy within the co-moving volume though the increase is never capable of reversing the redshifting due
to the expansion. After annihilation the electromagnetic temperature $T_{\gamma}$ is related to the neutrino
temperature $T_{\nu}$ via the well known $T^{3}_{\nu}/T^{3}_{\gamma} = 4/11$ and the time-electromagnetic temperature
relationship evolves to become
\begin{equation}
t\,T_{\gamma}^{2} = 1.32\;{\rm MeV^{2}\,s} \label{eq:tpost}
\end{equation}
again with the standard assumption of zero neutrino degeneracy. The changing time-temperature relationship is important
here because it determines how much neutron decay can occur before nuclei begin to form. Hereafter we drop the
subscript for $T_{\gamma}$ so that whenever we mention temperature it is always the electromagnetic.

So far we have made no mention of the complex nuclei. The temperatures are so high that their abundances are suppressed
relative to the free nucleons but of course the nuclear reactions that form them do occur. In contrast with the
$\n\leftrightarrow \p$ interconversion processes the nuclear reactions, such as $\n+\p \leftrightarrow \D+\gamma$, are
rapid at $T_{np}$ and so the abundances reach, and maintain, chemical/nuclear statistical equilibrium (NSE). In
equilibrium the abundance\footnote{The term `abundance' is also used for the ratio $Y_{A}/Y_{H}$}, $Y_{A}=n_{A}/n_{B}$,
of the complex nuclei $A$ is derived from $\mu_{A} = Z\,\mu_{\p} + (A-Z)\,\mu_{\n}$ so inserting the expressions for
the Boltzmann number density we find.
\begin{equation}
Y_{A} = \frac{g_{A}\,A^{3/2}}{2^{A}}\, \left[ n_{B}\, \left(\frac{2\,\pi}{m_{N}\,T}\right)^{3/2} \right]^{A-1}
\,Y_{\p}^{Z}\,Y_{\n}^{A-Z}\, e^{B_{A}/T}. \label{eq:NSE}
\end{equation}
From this equation we see that for a temperature of $T \sim 1\;{\rm MeV}$ the abundance of deuterons is $Y_{\D} \sim
10^{-12}$. We can rewrite equation (\ref{eq:NSE}) to illustrate what is happening in this temperature regime, by
replacing the thermal factors with the $Y_{\D}$ so that
\begin{eqnarray}
Y_{A} & = & \frac{g_{A}\,A^{3/2}}{2\,[3\,\sqrt{2}]^{A-1}}\,Y_{\p}^{1+Z-A}\,Y_{\n}^{1-Z}\,Y_{\D}^{A-1} \nonumber \\
 & & \times \exp \left( \frac{B_{A}-(A-1)B_{\D}}{T} \right).
\end{eqnarray}
This equation makes it much clearer that the abundance of nucleus with mass $A+1$ relative to nucleus with mass $A$ is
smaller by approximately $Y_{\D}$ indicating just how severe the suppression is when $Y_{\D}$ is small.

There are many different reactions in which the nuclei participate. For BBN the most important are \cite{Smith:1992yy}
\begin{eqnarray}
\n+\p & \leftrightarrow & \gamma +\D \label{eq:npgd}
\end{eqnarray}
\begin{subequations}
\begin{eqnarray}
\D+\D & \leftrightarrow & \n + \He[3] \label{eq:D+D->He3}\\
\D+\D & \leftrightarrow & \p + \T \label{eq:D+D->T} \\
\D+\p & \leftrightarrow & \gamma + \He[3] \label{eq:D+p}
\end{eqnarray}
\end{subequations}
\begin{subequations}
\begin{eqnarray}
\T+\D & \leftrightarrow & \n + \He \label{eq:T+D}\\
\T+\He & \leftrightarrow & \gamma +\Li
\end{eqnarray}
\end{subequations}
\begin{subequations}
\begin{eqnarray}
\He[3]+\n & \leftrightarrow & \p +\T \\
\He[3]+\D & \leftrightarrow & \p + \He \label{eq:He3+D}\\
\He[3]+\He & \leftrightarrow & \gamma + \Be
\end{eqnarray}
\end{subequations}
\begin{eqnarray}
\Li+\p & \leftrightarrow & 2\;\He \\
\Be+\n & \leftrightarrow & \p +\Li \label{eq:benpli}
\end{eqnarray}

In every case the rate of change of abundance of a nucleus $j$, a member\footnote{The set does not include non-nuclear
particles} of the set of products $\{P\}$, from any given reaction involving the reactant set $\{R\}$ is
\begin{equation}
\frac{1}{\nu_{j}}\,\frac{dY_{j}}{dt}= \Gamma_{R\rightarrow P} \prod_{i\,\in\{R\}} \frac{Y^{\nu_{i}}_{i}}{\nu_{i}!}
\label{eq:dydt}
\end{equation}
where the $\nu$'s are the stoichiometric coefficients and $\Gamma_{R\rightarrow P}$ is the rate of the reaction per unit
abundance of the reactants. In equilibrium, the production and destruction of the nuclei by any one reaction are almost
equal with only a small difference between them. This allows us to relate the forward, $\Gamma_{R\rightarrow P}$, and
back reaction, $\Gamma_{P\rightarrow R}$, rate coefficients i.e.
\begin{equation}
\Gamma_{R\rightarrow P} \prod_{i\,\in\{R\}} \frac{Y^{\nu_{i}}_{i}}{\nu_{i}!} \approx \Gamma_{P\rightarrow R}
\prod_{j\,\in\{P\}} \frac{Y^{\nu_{j}}_{j}}{\nu_{j}!}
\end{equation}
If we insert into this expression the Bolztmann distribution of the number densities then we find
\begin{eqnarray}
\Gamma_{P\rightarrow R} & \propto & \Gamma_{R\rightarrow P}\,\exp\left( \frac{(M_{P}-M_{R})- (\mu_{P}-\mu_{R})}{T} \right)\\
& \propto& \Gamma_{R\rightarrow P}\,\exp\left(\frac{-Q}{T}\right) \label{eq:gamma inverse}
\end{eqnarray}
where $M_{R}$ and $M_{P}$ ($\mu_{P}$ and $\mu_{R}$)are the sum of the reactant or product particle masses (chemical
potentials) and $Q$ is the difference. If the reaction involves only nucleons/nuceli and photons then $\mu_{P}=\mu_{R}$
and the `$Q$ value' is simply the difference in the total binding energy of reactants and products, but in those cases
where the reaction involves particles we have not included in $\{R\}$ and $\{P\}$ the equality is not ensured. An
analytic discussion of the effects BBN from the point of view of the reaction rates can be found in
\cite{Esmailzadeh:1990hf}.

As the Universe cools the reactions are unable to process the nuclei leading to a departure of the abundance of each
from the NSE abundance: the heavier nuclei departing earlier than the lighter.
\begin{figure}[htbp]
\begin{center}
\epsfxsize=3.4in \epsfbox{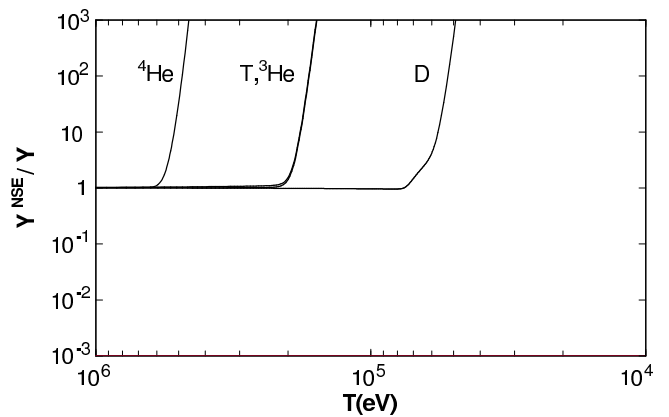} \caption{The ratio $Y_{i}^{NSE}/Y_{i}$, the NSE abundance to the actual abundance, of
\D, \T, \He[3] and \He~ as a function of the photon temperature when the baryon/photon ratio is $5.6 \times 10^{-10}$.
\label{fig:NSE departure} }
\end{center}
\end{figure}
The ratio of the NSE abundance and the actual abundance from a standard BBN code is shown in Fig. (\ref{fig:NSE
departure}) and was first discussed in Smith, Kawano \& Malaney \cite{Smith:1992yy}. The figure shows that the \He~
abundance departs from NSE when $T \sim 600\;{\rm keV}$ while \He[3] and \T~ depart at $T \sim 200\;{\rm keV}$. Below
$T \sim 200\;{\rm keV}$ only the abundance of the deuteron is given by its NSE value, the rest are many orders of
magnitude smaller. The departure from NSE is not because the rates fall beneath the Hubble rate but rather the
departure occurs because the amount that can be provided from the reactions falls short of the amount required to
\emph{remain} in equilibrium. For any particular nuclear species the former is simply the sum of all the relevant
production and destruction reactions listed in (\ref{eq:npgd}) through (\ref{eq:benpli}) while the latter can be found
from equation (\ref{eq:NSE}) (since deuterons are only beginning to form, we ignore the change in $Y_{\p}$ and $Y_{\n}$
and remembering $n_{B} \propto T^{3}$),
\begin{equation}
\frac{dY_{A}}{dt} = \frac{Y_{A}}{T}\,\frac{dT}{dt}\,\left[ \frac{3\,(A-1)}{2} -\frac{B_{A}}{T} \right].
\label{eq:dyNSE/dt}
\end{equation}

The departure from NSE has a major impact upon the reactions. The rate of change of nucleus $k$ in, for example, a
two-body reaction $i+j\leftrightarrow k+l$ with $i\neq j$, $k\neq l$, is
\begin{eqnarray}
\frac{dY_{k}}{dt} & = & Y_{i}\,Y_{j}\,\Gamma_{ij\rightarrow kl} - Y_{k}\,Y_{l}\,\Gamma_{kl\rightarrow ij} \\
 & = & Y_{i}\,Y_{j}\,\Gamma_{ij\rightarrow kl}\,
 \left[ 1 - \frac{Y_{k}}{Y^{NSE}_{k}}\,\frac{\,Y_{l}}{Y^{NSE}_{l}} \right],
\end{eqnarray}
with the superscript NSE indicating the equilibrium abundance, is now very lopsided because $Y_{k}/Y_{k}^{NSE}
\ll 1$. Essentially there is no destruction of $k$ via this reaction, the back reaction has switched off.

By $T \sim 200\;{\rm keV}$ the only compound nucleus in NSE is the deuteron. Its abundance is rising rapidly and BBN is
said to begin when the nuclear reactions finally begin to process the deuterium leading to the final departure from
NSE. The increasing significance of such reactions as $\D+\D\rightarrow \T+\p$ halts the rise in the deuterium
abundance and Bernstein, Brown \& Feinberg define the temperature at which BBN begins, $T_{BBN}$, as the point where
the deuteron abundance reaches its peak i.e. $dY_{\D}/dt = 0$. Using this definition they find $T_{BBN} \approx
B_{\D}/26 = 86\;{\rm keV}$, or $t_{BBN}=180\;{\rm s}$.

The second stage to BBN, from $1.2\;{\rm MeV}$ to $86\;{\rm keV}$, is characterized by departure from equilibria. At
its inception non-equilibrium is limited to the neutron/proton ratio: by its end all other nuclei have fallen out of
equilibrium with deuterium the last to succumb. Formation of the heavier nuclei in significant amounts begins to occur
towards the end of this stage and indeed the abundance of \He~ can already be larger than \D~ by $T_{BBN}$. What
$T_{BBN}$ represents is the point where the (leaky) dam bursts so to speak. The initiation of the next stage of BBN
proper is controlled by the deuteron binding energy $B_{\D}$ and therefore, in conjunction with $\delta m_{np}$, the
initial conditions for BBN proper are essentially controlled by just these two parameters.


\subsection{BBN proper}

From the departure of \n/\p~ weak equilibrium at $T \sim 1\;{\rm MeV}$ to the inception of BBN at $T \sim 100\;{\rm
keV}$ some of the free neutrons have decayed and reduced the pool available to be assimilated. But once the balance has
been tipped in favor of complex nuclei the free nucleons are rapidly dragged into and through $A=2$ and neutrons become
stabilized. This phase is BBN proper and during it the abundances of the complex nuclei can become very large as the
nuclear reactions process them. But by the time the Universe has cooled to the point where the reactions have ceased
virtually all the neutrons present at $T_{BBN}$ now reside in helium-4 with a small fraction in the trace abundances of
the other nuclei. The trace abundances of the intermediary nuclei \D, \T \, and \He[3] are the ashes of processing from
free nucleons to \He~. Because helium-4 essentially acts as the end-point of the reactions its abundance is very
insensitive to the exact details of the nuclear reactions that lead to its formation. The abundance of \He~ instead
probes the much earlier epoch of n/p freeze-out and the time delay until BBN commences. In contrast, the final
abundance of the intermediary nuclei \D, \T \,and \He[3] are strong functions of the nuclear physics. The efficiency of
the processing from free nucleons to \He~ depends crucially on the conditions during BBN, the temperature and the
interaction cross sections, so their abundances can vary markedly if BBN begins at a higher temperature or the cross
sections change.


\section{Nuclear Reaction Rates and $\Lambda_{QCD}$}
\label{sec:nuclear}

As discussed above, the deuteron binding energy and the neutron-proton mass difference are two of the most important
parameters that control Big Bang Nucleosynthesis. Each of these depends on the QCD scale, although each in a different
way.

Shifts in the QCD scale might not, at first glance, be expected to have significant effects.  This is because most
nuclear quantities, e.g. nuclear masses, nucleon masses, will shift together with this scale and therefore not be
observable. However, there is no reason to expect that all the fundamental constants shift at the same rate, and the
relative change may cause observable effects. For example, while the masses of the proton and neutron are mainly
determined by the QCD scale, the difference between the neutron and proton masses is dominated by electroweak effects.
As a consequence, the ratio $(m_\n-m_\p)/m_\n$ may be sensitive to changes in the fundamental couplings. Furthermore
the pion mass will not shift in the same way as the nucleon or nuclear masses, and the pion mass is an important
ingredient in setting the deuteron binding energy.

The neutron-proton mass difference is approximately given by \cite{Gasser:1982ap}
\begin{equation}
\delta m_{np} = M_\n - M_\p = m_d - m_u - \alpha M_{elm}
\end{equation}
The coefficient of $\alpha$ in the electromagnetic contribution, $M_{elm}$, is determined by strong interactions and
therefore proportional to $\Lambda_{QCD}$.  We use the estimate  $\alpha M_{elm}\simeq 0.76 \, {\rm MeV}$, while the
difference in the down and up quark masses is approximately 2 MeV \cite{Gasser:1982ap}. Other calculations of $\alpha
M_{elm}$ yield different results, for example \cite{Henley:1990ku}. Shifts in the mass of the up and down quark masses
depend on the vacuum expectation value of the Higgs and their respective Yukawa couplings. Specifically,
\begin{eqnarray}
\label{eq:shiftdmnp} \Delta ( \delta m_{np}) & = & - \left( {\Delta \alpha \over \alpha} + {\Delta \Lambda_{QCD} \over
\Lambda_{QCD}} \right) \alpha M_{elm} \nonumber \\ & & + \left( {\Delta y_d - \Delta y_u \over y_d - y_u} + {\Delta v
\over v} \right) \, (m_d - m_u)
\end{eqnarray}
Here the $y$'s are the Yukawa couplings and $v$ is the vacuum expectation value (vev) of the Higgs boson. From this
equation, it can be seen that the shift in the neutron-proton mass difference is determined by several fundamental
parameters, not just $\Lambda_{QCD}$.

In principle, we should define a scale which remains fixed, while others vary.  This could be a scale such as the GUT
scale.  In our calculations in the next section however, we only vary $\Lambda_{QCD}$ so this scale could even be the
electron mass. A discussion of scales as well as a preliminary estimate of the deuteron binding energy may be found in
\cite{Dent:2001ga}

The scale $\Lambda_{QCD}$ also has great importance in determining the binding energy of the deuteron.  The small
binding energy prevents a significant abundance of deuterium from building up early in big bang nucleosynthesis. And
therefore, as discussed above, the deuteron binding energy controls the nuclear flow to elements with greater mass
number. The long-range part of the nuclear force is governed by one-pion exchange and therefore the deuteron binding
energy is sensitive to pion properties. The mass of the pion is determined by the Gell-Mann-Oakes-Renner relation,
\cite{gmor}
\begin{equation}
f^{2}_\pi m_\pi^2 = (m_u + m_d) < \bar{q} q>
\end{equation}
Here $f_\pi$ is the coupling of the pion to the axial current, and $<\bar{q} q>$ is the quark condensate.

Shifts in the pion mass are then calculated in a similar manner to Eq (\ref{eq:shiftdmnp}) and determined by the shifts
the Yukawa couplings, v and $\Lambda_{QCD}$. Since $f_\pi \propto \Lambda_{QCD}$ and $< \bar{q} q> \propto
\Lambda_{QCD}^3$,
\begin{equation}
{\Delta m_\pi \over m_\pi} = {1 \over 2} \left[ {\Delta \Lambda_{QCD} \over \Lambda_{QCD}} +
{\Delta v \over v} + {\Delta y_u + \Delta y_d  \over y_u + y_d} \right] \label{eq:shiftmpi}
\end{equation}
Recent studies of the dependence of the nuclear potential on the mass of the pion have shown a strong variation with
the mass of the pion \cite{Beane:2002vq, Beane:2002xf, Epelbaum:2002gb}. According to these authors, changes in the
pion mass of a few percent could lead to changes in the deuteron binding energy of a factor of two. Although a
relationship between the pion mass and the deuteron binding energy exists, \cite{Beane:2002vq,
Beane:2002xf,Epelbaum:2002gb}, several parameters in this relationship are uncertain.  These include parameters that
describe the pion nucleon coupling and the coefficient of a quark mass dependent four nucleon operator.

Next we turn to the reaction rates themselves that are most important for determining the deuterium and helium
abundances.

The first is the neutron-proton interconversion reaction listed in (\ref{eq:ndecay}) through (\ref{eq:nnue}). The rates
for the reactions must include the temperature chemical potentials of the electrons and neutrinos which also
participate in the reactions. In the Born approximation the neutron/proton interconversion rates are
\begin{widetext}
\begin{eqnarray}
\Gamma_{\n\rightarrow \p} & = & \frac{G_{F}^{2}\,
  (C_{V}^{2}+3\,C_{A}^{2})}{2\,\pi^{3}} \int_{m_{e}}^{\infty}
   dE_{e}\,E_{e}\,p_{e}
\, \big[ (E_{e}+\delta m_{np})^{2}\,
    f_{\bar{e}}(E_{e})\,\bar{f}_{\bar{\nu}}(E_{e}+\delta m_{np})
      + (E_{e}-\delta m_{np})^{2}\,\bar{f}_{e}(E_{e})\,f_{\nu}
       (E_{e}-\delta m_{np}) \big] \nonumber \\ & &
 \label{eq:Gamma n}\\
\Gamma_{\p\rightarrow \n} & = & \frac{G_{F}^{2}\,
   (C_{V}^{2}+3\,C_{A}^{2})}{2\,\pi^{3}} \int_{m_{e}}^{\infty}
    dE_{e}\,E_{e}\,p_{e}
 \big[ (E_{e}+\delta m_{np})^{2}\,
    \bar{f}_{\bar{e}}(E_{e})\,f_{\bar{\nu}}(E_{e}+\delta m_{np})
   + (E_{e}-\delta m_{np})^{2}\,f_{e}(E_{e})\,\bar{f}_{\nu}
   (E_{e}-\delta m_{np}) \big] \nonumber \\ & &
 \label{eq:Gamma p}
\end{eqnarray}
\end{widetext}
where $f(E)$ is the Fermi-Dirac distribution, $\bar{f}(E)$ its compliment and the subscripts indicate the chemical
potential and temperature to be used in $f$. Though these two expressions capture the essence of how $\delta m_{np}$
enters into these rates there are a number of corrections that must be taken into account before the $\n
\leftrightarrow \p$ reaction rates reach sufficient accuracy \cite{discus,wilkinson,esposito} given the accuracy of the
data we will eventually use in our constraints.

The next is the reaction which creates deuterium, $\n + \p \rightarrow \D + \gamma$, which is well studied in a low
energy effective field theory without pions \cite{Chen:1999bg,Rupak:1999rk}. These authors give an expression for the
cross section as a function of the deuteron binding energy, the scattering length in the singlet channel, the phase
shift, and so on. In principle each of these parameters should be treated as free, however, the one which has by far
the largest leverage on the rate is the binding energy of the deuteron.  In our calculation, we use the expression for
the cross section given by \cite{Rupak:1999rk} and integrate over the thermal distribution of the particles in the
manner prescribed in Fowler, Caughlan and Zimmerman \cite{FCZ1967}.

The other crucial reaction is the one that destroys most of the deuterium, $\D + \D \rightarrow \T + \p$ during BBN
proper. Ideally, we would like to use the results of a nuclear effective field theory calculation in order to determine
the dependence of this rate on the pion mass, but there are no effective field theory calculations available yet.


\section{Non-Standard BBN}
\label{sec:nonstand}

Standard BBN is well studied problem and even though the compatibility of its predictions with the observed abundances
remains a contentious issue much of the recent efforts have been towards using it as a test of the state of the
Universe at the earliest epochs. Modifying standard BBN to include new effects can be relatively painless: for example,
the use of BBN as a probe of the number of light neutrino species or the chemical potentials of the $\mu$ and/or $\tau$
neutrino flavors is simply a case of modifying the energy density and hence the expansion rate with no direct influence
upon the nuclear physics \cite{SSG1977,YSSR1979,KSSW2001}. Similarly the expansion rate can be modified by changing the
cosmological equations \cite{CSS2001,KS2003}, introducing extra energy density in the form of `Quintessence'
\cite{SDT1993,SDTY1992,BHM2001,KS2003} and even quite general evolutions have been explored \cite{CK2002}. The effects
of  new physics from scenarios such as a non-zero electron neutrino chemical potential
\cite{BY1977,KS1992,KKS1997,OSTW1991,S1983,WFH1967,YB1976,Betal03} or neutrino oscillations/mass/decay
\cite{F2000,KSW2000,K1992,KS1982,SSF1993,SFA1999}, show up primarily in the neutron-proton interconversion reactions.
There the known expressions for these reaction rates permit a high degree of confidence that all the consequences of
the new physics have been accurately taken into account.

By far the most difficult proposals to implement touch every nuclear reaction. For example, the impact of a varying
fine structure constant was examined by Bergstr\"{o}m, Iguri and Rubinstein \cite{BIR1999}, later improved upon by
Nollett and Lopez \cite{NL2002}, and to examine its effects these authors had to recalculate every relevant rate. In
this type of study one is forced to make a number of approximations because we do not have a complete understanding of
how a change in a fundamental constant alters low energy nuclear physics parameters such as the binding energies and
cross sections. An alternative approach would be to map the parameter space from a single, or handful of, fundamental
constant(s) to a greater number of nuclear physics parameters, and then constrain the nuclear physics parameters, while
treating them as independent. This way we avoid the uncertainty in how changes in the fundamental constants manifest
themselves in nuclear quantities but this comes at the expense of exploring larger regions of parameter space. Despite
this trade-off, we shall turn in this direction.

The number of nuclear physics parameters is huge but, as we have been at pains to stress, the most important two are
$\delta m_{np}$ and $B_{\D}$ and indeed we can limit our parameter space to just these two quantities. Before we set
out our plan for constraining $\delta m_{np}$ and $B_{\D}$ we examine why we can restrict the number of nuclear
parameters to just this limited set yet still obtain good predictions for the results of BBN and how these two
parameters will affect the predicted primordial abundances.


\subsection{Simplifying BBN}

A prediction for the mass fraction of helium-4 is reliably calculated by simply counting the number of neutrons that
become stabilized inside nuclei. This approximation for the helium-4 mass fraction emerges from the empirical results
of standard BBN discussed above where we showed that all the neutrons that survive BBN reside in helium-4. To determine
the number of stabilized neutrons we must first be able to calculate the neutron-proton ratio during the earliest,
weak-equilibrium, phase of BBN and then, secondly, follow it through to the inception of BBN proper. The first is
relatively easy to implement since we have known expressions to use for the neutron-proton interconversion rates. As we
mentioned earlier, there are many corrections to equations (\ref{eq:Gamma n}) and (\ref{eq:Gamma p}) that must be
included in order to improve their accuracy. Of these we only explicitly included the radiative electromagnetic
corrections and do not introduce the corrections for the finite mass or thermal radiative corrections. Since they are
relatively small when compared with the effects we want to consider, we take them into account only by an overall scale
factor normalized to the measured neutron lifetime. The second step means that we must be able to follow the neutrons
into the first reaction that must occur in processing the nuclei: $\n+\p \leftrightarrow \D+\gamma$. As we mentioned in
section \S\ref{sec:nuclear}, an analytic expression for the cross section of this reaction is available from Rupak
\cite{Rupak:1999rk} which shows how the deuteron binding energy enters into this important quantity.

As we learned from standard BBN the final abundance of \He~ is very insensitive to the exact rate of its formation so
if we truncated our reaction network at this step we could still get a good estimate for the helium-4 mass fraction and
thus we only need two parameters for the calculation: the neutron-proton mass difference $\delta m_{np}$ and the
deuteron binding energy $B_{\D}$. But with two parameters (three if we include the baryon to photon ratio $\eta$) and
one prediction a degeneracy is expected to arise between $\delta m_{np}$ and $B_{\D}$ in the abundance of \He~.
Therefore, it is worth endeavoring to make a prediction for the abundance of at least one other nucleus. The most
obvious candidate is deuterium because its primordial abundance is also a function of these same two parameters. To
make a prediction for the amount of unprocessed deuterium we have to follow not only how this nucleus is formed
courtesy of the $\n+\p \leftrightarrow \D+\gamma$ reaction but also how much is destroyed. The majority of the
destruction of deuterium occurs via the reactions (\ref{eq:D+D->He3}), (\ref{eq:D+D->T}) and (\ref{eq:T+D}) with
(\ref{eq:D+p}) and (\ref{eq:He3+D}) following behind in importance \cite{Smith:1992yy}. As we saw in equation
(\ref{eq:gamma inverse}) the inverse rate is related to the forward rate by a term that is proportional to $\exp(-Q/T)$
and $Q$ may be expressed in terms of the binding energies. By including these reactions we are adding the tritium and
helium-3 binding energies explicitly to the $\delta m_{np}$ and $B_{\D}$ parameter list through this exponential
dependence.  (In addition, there may be additional dependence of these binding energies in the forward cross sections,
although there is no explicit calculation currently available.) But, here another result from standard BBN comes to our
rescue: the nuclear abundances of all the nuclei heavier than \D~ during BBN fall far below their nuclear statistical
equilibrium values and the inverse reactions essentially switch off long before BBN begins. This allows us to use
tritium and helium-3 as the substitutes for helium-4 \emph{if} we can assume that any neutron that ever forms one of
these nuclei is prevented from ever participating in another reaction. If we are able to ignore all inverse rates then
we are eliminating the explicit dependence upon the binding energies of tritium and helium-3 (although the cross
sections will still implicitly be functions of these quantities). Furthermore, if we truncate our reaction network at
these two $A=3$ nuclei then we need only be concerned with reactions (\ref{eq:D+D->He3}) and (\ref{eq:D+D->T}). With
the absence of an analytic expression to use, we take the cross sections of these two reactions to be to be related to
the size of the deuteron radius i.e. $\sigma \sim 1/(m_{N}\,B_{\D})$. In terminating the network so early we miss the
significant destruction of deuterium from $\T+\D \rightarrow \n + \He$ and the less important $\He[3]+\D \rightarrow \p
+\He$ but nevertheless, we can use the abundance of \D~ from the calculation as an estimate of the primordial abundance
even though this approximation is much cruder than the one used for the primordial \He~ abundance.

Our approximation to BBN obviously introduces uncertainty into its predictions but the agreement with standard BBN
calculations is remarkably impressive. As a function of the baryon to photon ratio $\eta$ the helium-4 abundance is
systematically too low by $\lesssim$ 1\% compared to a standard calculation. The reason lies in the fact mentioned
previously: the abundance of \He[4] surpasses the abundance of \D~ at $T_{BBN}$, the beginning of BBN proper, but since
we have essentially removed this nucleus from the network we overestimate the amount of neutron decay and therefore
calculate a lower than expected yield of helium-4. In contrast deuterium is systematically overestimated but this time
by $\sim$ 50\%. Neglecting $\T+\D \rightarrow \n + \He$ has introduced insufficient deuterium destruction ,which is of
order the observation uncertainty. We will show later how powerful the use of deuterium could be if we had more
information about how other nuclear cross sections/binding energies vary with the pion mass.


\subsection{The effect of changing $B_{\D}$}

\begin{figure}[htbp]
\begin{center}
\epsfxsize=3.4in \epsfbox{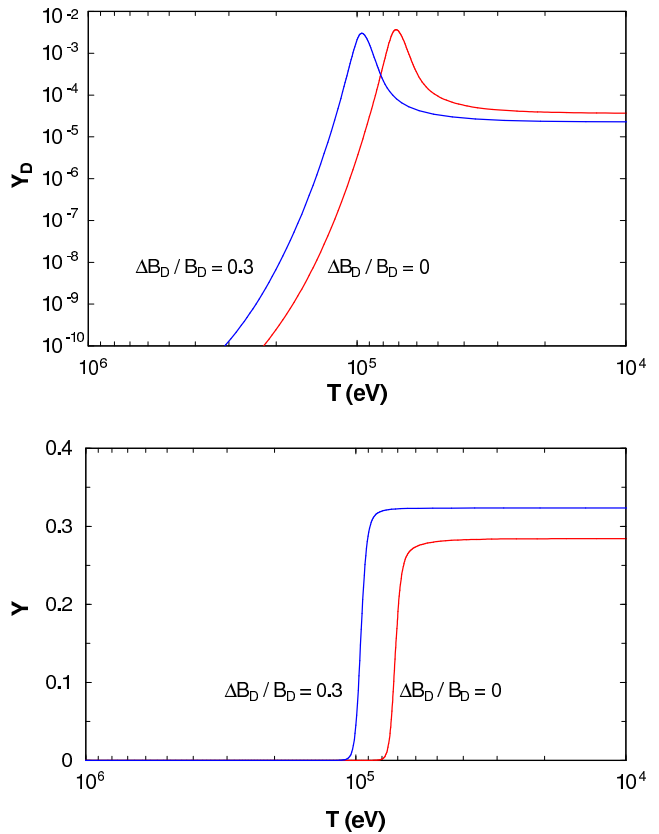} \caption{The deuterium abundance, $Y_{\D}$, top panel, and the helium-4 mass
fraction, $Y$, bottom panel, as functions of the photon temperature for $\Delta B_{\D}/B_{\D} =0$ and $\Delta
B_{\D}/B_{\D} =0.3$ when the baryon/photon ratio is $6.14 \times 10^{-10}$. \label{fig:vsbd} }
\end{center}
\end{figure}

The effects of a change in $B_{\D}$ occur through both the change in the NSE deuteron abundance and the cross sections
that process \D~ to heavier nuclei. As $B_{\D}$ increases the deuteron becomes more stable and consequently more
difficult to disassociate. Therefore we expect as $B_{\D}$ increases BBN will commence at a higher temperature/earlier
epoch. Perhaps counterintuitively the increased stability of the deuteron leads to a decrease in the primordial
deuterium abundance. The rate coefficients are functions of both the temperature and the powers of the Baryon density
$n_{B}$ so if $T_{BBN}$ increases interactions occur more rapidly even if the cross sections have changed. This leads
to a more efficient processing of the neutrons to \He~ so that the amount of unprocessed, and thus the primordial,
deuterium decreases. From a more efficient processing of the nucleons we would expect an increase in the primordial
\He~ abundance and this is enhanced by the shorter interval between n/p freeze-out and the beginning of BBN and
consequently less n-decay. These expectations are confirmed in figure (\ref{fig:vsbd}) where we plot the deuterium
abundance $Y_{\D}$ and helium-4 mass fraction, confusingly denoted also by $Y$, as functions of the temperature for
$\Delta B_{\D}/B_{\D} =0$ and $\Delta B_{\D}/B_{\D} =0.3$.
At temperatures above $100\;{\rm keV}$ when deuterium is in NSE the offset due to the change in its binding energy is
clear, as is the shift in the peak deuterium abundance, and hence $T_{BBN}$. The lower figure shows that the change in
the deuterium binding energy leads to an increase in the final helium-4 mass fraction as predicted.


\subsection{The effect of changing $\delta m_{np}$}

There are two effects that become apparent when we change the neutron-proton mass difference: an altered neutron
lifetime and a change in the n/p freeze-out ratio. Before we add the radiative corrections, the neutron lifetime
$\tau_{\n}$ is simply
\begin{eqnarray}
\label{eq:neutron} \frac{1}{\tau_{\n}} & \approx & \,\frac{G_{F}^{2}\,(C_{V}^{2}+3\,C_{A}^{2})}{2\,\pi^{3}} \nonumber \\
 & & \times \int_{m_{e}}^{\delta m_{np}} dE_{e}\,E_{e}\,p_{e}(E_{e}-\delta m_{np})^{2} \label{eq:n lifetime} \\
& = & \delta m_{np}^{5}\,I(m_{e}/\delta m_{np}). \label{eq:n Delta5}
\end{eqnarray}
Since function $I(m_{e}/\delta m_{np})$ varies less rapidly than $\delta m^{5}_{np}$ as $\delta m_{np}$ increases the
lifetime drops considerably. We therefore expect more neutron decay in the period from neutron/proton freeze-out until
the inception of BBN and consequently less \He~ and less \D.

Secondly, the neutron-proton mass difference $\delta m_{np}$ sets the neutron to proton ratio prior to the cessation of
the interconversion reactions. From equations (\ref{eq:Gamma n}) and (\ref{eq:Gamma p}) we can see that increasing
$\delta m_{np}$ will lead to increases in both rates per particle but this is more than overwhelmed by the decrease in
the equilibrium neutron abundance $F \propto \exp(-\delta m_{np}/T)$. The increase in $\delta m_{np}$ therefore also
leads to lower primordial \D~ and \He~ abundances because there are less neutrons around that can form these nuclei. We
can add some quantitativeness to these remarks by following the simplified treatment of neutron/proton freeze-out of
Bernstein, Brown \& Feinberg \cite{BBF1989}. By ignoring \n-decay\footnote{Although Bernstein, Brown \& Feinberg
\cite{BBF1989} ignore n-decay in deriving their expressions they still require neutrons to decay eventually in order to
normalize the interconversion reactions.} these authors derive an expression for the freeze-out abundance of neutrons
$Y^{\star}_{\n}$ as
\begin{widetext}
\begin{equation}
Y^{\star}_{\n} = \int_{0}^{\infty}\,\frac{dq}{2} \,\frac{1}{1+\cosh(q)}\,
\exp\left(-\frac{f(m_{e}/\delta m_{np})}{\tau\,\delta m_{np}^{2}}\,
\left[\left(\frac{4}{q^{3}}\,+\frac{3}{q^{2}}+\frac{1}{q}\right) + \left(\frac{4}{q^{3}}+\frac{1}{q^{2}} \right) e^{-q}
\right] \right) \label{eq:y freezeout}
\end{equation}
\end{widetext}
where the function $f(m_{e}/\delta m_{np})$ varies slowly at $m_{e}/\delta m_{np} \approx 1/2$. The change of
$Y^{\star}_{\n}$ in (\ref{eq:y freezeout}) is therefore primarily due to the effect of the $\tau\,\delta m_{np}^{2}$
factor in the denominator of the exponential. From equation (\ref{eq:n lifetime}) we know that the lifetime scales as
$\sim 1/\delta m_{np}^{5}$ so the factor in the exponent is proportional to $1/\delta m_{np}^{3}$. As $\delta m_{np}$
increases the exponential term in (\ref{eq:y freezeout}) fades more rapidly yielding a faster convergence with $q$ and
so confirming our expectation of a lower freeze-out abundance of neutrons.

Finally, as a consequence of the lower freeze-out neutron abundance the NSE deuteron abundance is lowered. In turn the
temperature, $T_{BBN}$, at which BBN proper commences is also lower because the reactions such as $\D+\D\rightarrow
\T+\p$ will not begin to process the deuterium until a slightly lower temperature. The Universe is therefore older at
the beginning of BBN thus permitting even more neutron decay than would be expected from just the change in neutron
lifetime. For \He~ this effect has the same sign as the previous two, as $\delta m_{np}$ increases the amount of \He~
decreases. In contrast, a lower $T_{BBN}$ actually leads to an increase in \D~ because there will be less destruction
of this nucleus but the increase is insufficient to compensate for the previous two effects.

\begin{figure}[htbp]
\begin{center}
\epsfxsize=3.4in \epsfbox{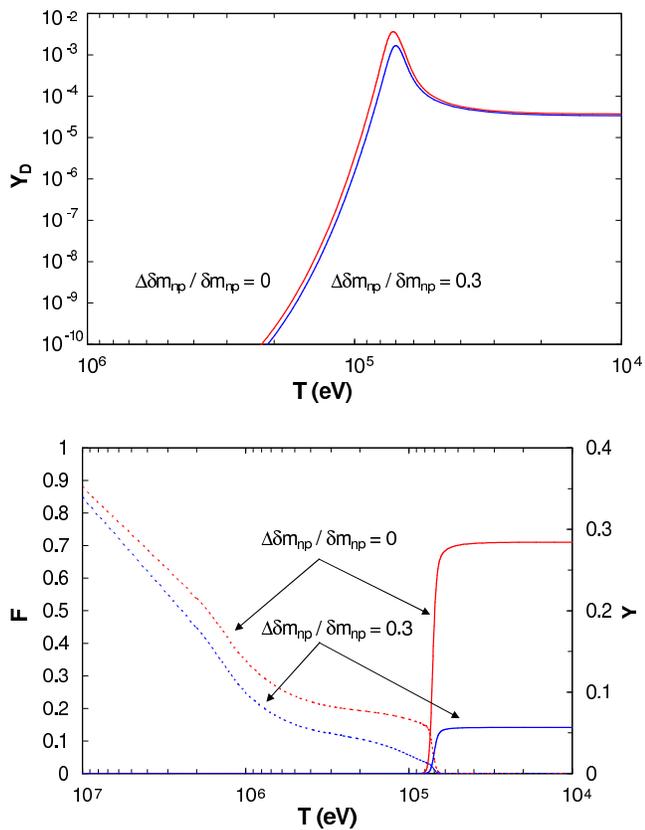} \caption{The deuterium abundance, $Y_{\D}$ (top panel), the (free) neutron/proton
ratio, $F$, (bottom panel, dashed lines) and the helium-4 mass fraction, $Y$, (bottom panel, solid lines) as functions
of the photon temperature when $\Delta \delta m_{np} /\delta m_{np} =0$ and $\Delta \delta m_{np} /\delta m_{np} =0.3$
with a baryon/photon ratio is $6.14 \times 10^{-10}$. \label{fig:vsdmnp} }
\end{center}
\end{figure}

The effects of a change in $\delta m _{np}$ are best summarized in figure (\ref{fig:vsdmnp}) where we plot the
deuterium abundance, $Y_{\D}$, the (free) neutron/proton ratio, $F$, and the helium-4 mass fraction, $Y$, as function
of the photon temperature at $\Delta \delta m_{np} /\delta m_{np} =0$ and $\Delta \delta m_{np} /\delta m_{np} =0.3$.
At high temperatures, above $\sim 1\;{\rm MeV}$, the shift in the neutron-proton mass difference is seen as the
difference in the two neutron/proton ratio curves while the change in neutron lifetime appears as the different
gradients for these curves close to $T \sim 10^{5}$. The different amounts of neutrons that survive from $T_{np}$ to
$T_{BBN}$ results in the large shift in the helium-4 mass fraction. This is primarily due to the change in the neutron
lifetime but is enhanced by a small shift in the onset of BBN proper as witnessed in the small shift in the position of
the peak deuterium abundance seen in the top panel.


\subsection{Constraining $\delta m_{np}$ And $B_{\D}$}

\begin{figure}[htbp]
\begin{center}
\epsfxsize=3.4in \epsfbox{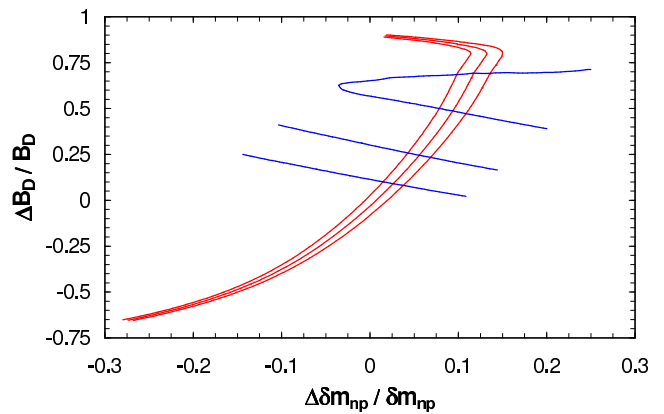} \caption{The iso-abundance contours $\D/{\rm H}$ of deuterium (horizontal)and
iso-mass fraction contours of helium-4 (vertical) from our calculation as a function of the fractional change in the
neutron-proton mass difference $\delta m_{np}$ and the deuteron binding energy $B_{\D}$ at $\eta = 6.14 \times
10^{-10}$. From top to bottom the deuterium contours are: $4 \times 10^{-5}, 3 \times 10^{-5}$ and $2 \times 10^{-5}$
while from left to right the helium-4 contours are $0.25, 0.24$ and $0.23$. \label{fig:isoabundances} }
\end{center}
\end{figure}

We can use the discussion in the last two sections to show how a degeneracy between $\delta m_{np}$ and $B_{\D}$ in the
prediction of the helium-4 mass fraction occurs and, furthermore, illustrate why using deuterium is/could be the ideal
foil. Raising $\delta m_{np}$ will lower the neutron/proton freeze-out ratio and so, in turn, reduce the helium-4 mass
fraction but if we also increase $B_{\D}$ then this can be entirely compensated by an earlier inception of BBN proper,
the point where neutrons become stabilized inside nuclei. At the same time the predicted abundance of deuterium is
expected to fall as we increase $\delta m_{np}$ but an increase in $B_{\D}$ also leads to less deuterium . Thus, while
the prediction for the abundance of deuterium also possesses a degeneracy between the two parameters, it is orthogonal
to that of helium-4. Therefore, using both allows us in principle, to gain strong constraints. This orthogonality is
best illustrated by figure (\ref{fig:isoabundances})
where the complementarity of the two nuclei is evident.

The figure also shows where the approximations begin to fail: significant deviations begin to occur when $\Delta B_{\D}
/B_{\D} \sim 0.7$, which, using the binding energy for the deuteron of $2.22\;{\rm MeV}$, is remarkably early. The
explanation lies in the Q-value of the reactions we have used. During the phase of NSE we require the binding energies
of tritium and helium-3 in order to compute the reverse reactions and for these we used the present measured values.
For reference the binding energies are $B_{\T} = 8.48\;{\rm MeV}$ and $B_{\He[3]} = 7.72\;{\rm MeV}$. The abundances of
tritium and helium-3 while they are in NSE are so small that even if this introduces a considerable error the final
results will not reflect this fault. The Q-value for the reaction $\D+\D \leftrightarrow \n + \He[3]$ is $Q =
B_{\He[3]} - 2 B_{\D}$ while that for $\D+\D \leftrightarrow \p + \T$ is $Q = B_{\T} - 2 B_{\D}$. So as we increase the
deuteron binding energy the Q-values for both reactions decrease and at a 70\% increase the Q-value for $\D+\D
\leftrightarrow \n + \He[3]$ reaches zero. With such a low Q-value the inverse reactions are significant and our use of
tritium and helium-3 as neutron sinks is no longer valid. In our numerical calculations, when the Q-values of these
reactions became small, we saw a very different flow of the nuclei through the reaction network compared to standard
BBN. If we persist with the increases in $B_{\D}$ and make the Q-values negative we enter very dangerous territory
since our simple rescaling of the cross-sections cannot still apply. Endothermic reactions are very different from
exothermic at low energy/tempertaure. This is not to say that the deuteron binding energy cannot be greater than 1.7
$B_{\D}$ during BBN, it is simply a statement that we cannot reliably predict the abundances in this domain. But a 70\%
increase in the deuteron binding energy cannot be regarded as a safe upper limit to the permitted variation in
$B_{\D}$. At 50\% the change in the deuteron binding energy is $\sim 1\;{\rm MeV}$. Since we do not know if the change
in $B_{\T}$ and $B_{\He[3]}$ is correlated or anti-correlated with $B_{\D}$, if $B_{\T}$ and $B_{\He[3]}$ vary with
similar magnitude to $B_{\D}$, the Q-value of $\D+\D \leftrightarrow \n + \He[3]$ may already have been forced to zero
at a 1 MeV increase in $B_{\D}$ and our approximations cannot be used. Therefore we estimate a 50\% increase in
$B_{\D}$ as a limit to the permitted variation of this parameter, in the absence of further information about how the
cross sections and binding energies vary with pion mass.

In contrast, it appears we can vary $\delta m_{np}$ in the region we want to explore without running into similar
difficulties. There is a lower limit to the variation we can allow: when $\delta m_{np}$ drops beneath the electron
mass, a decrease in $\delta m_{np}$ of 60\%, neutrons cannot decay so that our rescaling of the reaction rates with the
neutron lifetime to correct for the finite mass and thermal radiative corrections cannot be applied. Again, a
neutron-proton mass difference smaller than the electron mass during BBN cannot be ruled out, it is just that we cannot
calculate what occurs when this happens.

We are now in the position where we can begin to constrain our two parameters $\delta m_{np}$ and $B_{\D}$ by comparing
the predicted abundances with observation. The primordial abundance $\D/{\rm H}$ of deuterium is taken to be $\D/{\rm
H} = ( 2.6 \pm 0.4 )\times 10^{-5}$ \cite{Betal2003} while we use the Olive, Steigman and Walker \cite{OSW2000} value
of $Y = 0.238 \pm 0.005$ for the helium mass fraction $Y$. The exact primordial abundances remain a topic of debate
with two, largely incompatible, determinations for the helium mass fraction \cite{ITL1997,IT1998,OS1995,OSS1997,FO1998}
and excessive scatter in the measurements of deuterium \cite{Ketal2003,Betal2003} but these two nuclei still represent
the best probes of BBN because the other nuclei that could be used, such as \He[3] and \Li, suffer from large
uncertainties in the derivation of their primordial values. From comparing the observed abundances of \D~ and \He~ and
their associated errors with the iso-abundance contours in figure (\ref{fig:isoabundances}) it is apparent that
helium-4 will be the chief source of constraints on $\delta m_{np}$ while deuterium will play the same role for
$B_{\D}$. The errors for these observations also reflect the level to which we must beat down the systematic errors in
order to avoid contaminating our results with large offsets. For helium-4 we have succeeded handsomely since $0.005$
represents a 2\% error on $0.238$ while our systematic was at 1\%. For deuterium we have not done so well since the
observation has an error of 15\% and our systematic was at 50\%.

\begin{figure}[htbp]
\begin{center}
\epsfxsize=3.4in \epsfbox{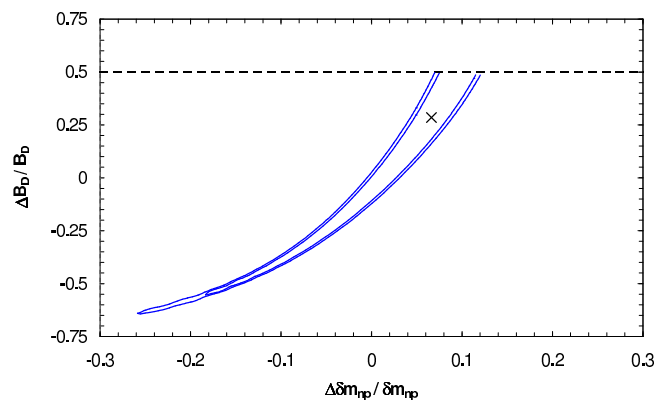} \caption{The 95\% and 99\% $\chi^{2}$ contours derived from using a primordial
deuterium abundance of $\D/{\rm H} = ( 2.6 \pm 0.4 ) \times 10^{-5}$ and a primordial helium-4 mass fraction of $Y =
0.238 \pm 0.005$. The cross, located at $\Delta \delta m_{np}/\delta m_{np} = 0.07$ and $\Delta B_{\D}/B_{\D} = 0.29$,
indicates the position of the best fit point. \label{fig:chi2 contours} }
\end{center}
\end{figure}

So in addition to the observational errors $\sigma_{\D}$ and $\sigma_{Y}$ we must also include into the analysis the
systematic errors, $S_{\D}$ and $S_{Y}$, we introduced when we made our approximations. The systematic errors differs
from the statistical observational errors in that they are correlated. The covariance matrix, $V$, is
therefore of the form
\begin{equation}
V = \begin{pmatrix} \sigma_{\D}^{2}+S_{\D}^{2} & S_{\D}S_{Y} \\ S_{\D}S_{Y} & \sigma_{Y}^{2}+S_{Y}^{2} \end{pmatrix}.
\end{equation}
If we knew the exact predictions for $Y_{\D}$ and $Y$ at any combination of $\delta m_{np}$, $B_{\D}$ and $\eta$ then
we could easily calculate $S_{\D}$ and $S_{Y}$ by comparing the exact value with that achieved with our approximations.
Unfortunately this is not the case and we can only make this comparison at $\Delta \delta m_{np}/\delta m_{np}=\Delta
B_{\D}/B_{\D}=0$. From comparing our results with that from a standard BBN code we can represent the systematic errors
as the product of the fractional errors and the predicted values.  We take the fractional errors to be the same at
arbitrary $\delta m_{np}$ and $B_{\D}$. With this understanding we can approximate the covariance matrix at all $\delta
m_{np}$ and $B_{\D}$ and calculate a likelihood via
\begin{equation}
{\cal L} = \frac{1}{2\pi \sqrt{|V|}}\,\exp\left(\frac{-{\bf \delta^{T}}\,V^{-1}\,{\bf \delta}}{2} \right)
\end{equation}
where the vector ${\bf \delta}$ is ${\bf \delta} = \{ Y_{\D}-\hat{Y}_{\D}, Y-\hat{Y}\}$ and the hat indicates the
observed values. If we also allow $\eta$ to vary then we have three adjustable parameters and only two constraints so
we choose to fix this quantity at the value of $\eta = 6.14 \times 10^{-10}$ given by the WMAP observations
\cite{Setal2003}. The CMB, which is essentially `atomic physics meets cosmology', is not expected to show any
dependence upon $\Lambda_{QCD}$ but other fundamental constants relevant to atomic physics have been constrained by
using it \cite{Yoo:2002vw,KST1999,H1999}. At this fixed value we derive the constraints on $\delta m_{np}$ and $B_{\D}$
shown in figure (\ref{fig:chi2 contours}).
The figure shows that $\delta m_{np}$ and $B_{\D}$ are not as well constrained as we might hope because of the large
systematic uncertainty in the prediction for deuterium but nevertheless the figure clearly indicates that the
primordial abundances are only compatible along a narrow band in the $\delta m_{np}$ - $B_{\D}$ plane.

\begin{figure}[htbp]
\begin{center}
\epsfxsize=3.4in \epsfbox{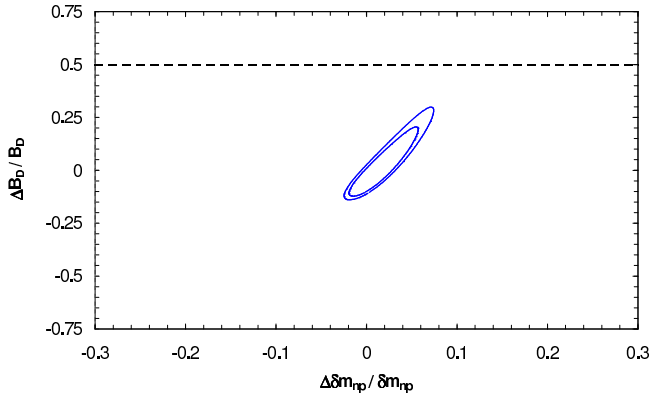} \caption{The 95\% and 99\% $\chi^{2}$ contours derived from using a primordial
deuterium abundance of $\D/{\rm H} = (2.6 \pm 0.4) \times 10^{-5}$ and a primordial helium-4 mass fraction of $Y =
0.238 \pm 0.005$ after rescaling the deuterium and helium-4 results from our numerical calculation by the systematic
errors we identified from comparison with a standard BBN code. This methodology is expected to produce results that are
representative of the situation where we do not have to terminate the nuclear reaction network at tritium and helium-3.
\label{fig:no systematic} }
\end{center}
\end{figure}

It is worthwhile exercise to show how much stronger the constraints would be if deuterium were not contaminated in this
way. To this end we show in figure (\ref{fig:no systematic}) the 95\% and 99\% $\chi^{2}$ contours when we rescale the
deuterium down by 50\% and helium up by 1\% and remove the systematic errors from the covariance matrix. In this way we
can simulate the situation of a complete understanding of how the cross sections and the binding energies are related
and there were no reason to truncate the reaction network. The size of the contours is now determined by the errors in
the observational values we used and we can see that the current deuterium abundance, which primarily constrains
$B_{\D}$ would permit only a variation of 20\% in this parameter at 95\% confidence while the neutron-proton mass
difference would be constrained to within 4\%, again at 95\% confidence. BBN can provide meaningful constraints on the
extent to which $\delta m_{np}$ and $B_{\D}$ could differ from their current values at the earliest epochs of the
Universe if we could determine, with more reliability, how the nuclear data we need for the calculation depends upon
either these two parameters or on the underlying fundamental constants.
Finally, we show how we can begin to remove portions of the $\delta m_{np}$ - $B_{\D}$ parameter space by using their
relationships with the fundamental constants such as $\Lambda_{QCD}$.
We would like to use the relationship for the deuteron binding energy as a function of the pion mass derived in Beane
and Savage \cite{Beane:2002vq,Beane:2002xf} and compare it with our constraints. Beane and Savage quote their results
in terms of the pion mass, but they are varying the ratio of the quark mass to $\Lambda_{QCD}$.  They show a range of
$B_D$ vs. $m_{\pi}$ which is effectively a function $B_D = \Lambda_{QCD} f(\sqrt{m_q/\Lambda_{QCD}})$, where the
function contains unknown, but constrained coefficients.  We use this function to vary the QCD scale and relate it to
the neutron-proton mass difference given in equation \ref{eq:shiftdmnp}, i.e.
\begin{equation}
{\Delta B_D \over B_D} = {\Delta \Lambda_{QCD} \over \Lambda_{QCD}}
+ {\Delta f \over f}
\end{equation}

\begin{figure}[htbp]
\begin{center}
\epsfxsize=3.4in \epsfbox{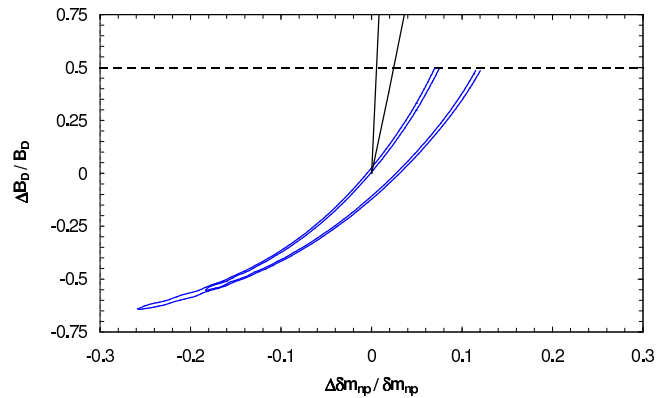} \caption{The 95\% and 99\% $\chi^{2}$ contours as in figure (\ref{fig:chi2 contours})
plus the limits upon $B_{\D}$ from \cite{Beane:2002vq,Beane:2002xf} after changing the variable from the pion mass to
the neutron-proton mass difference.
 \label{fig:chi2 with silas} }
\end{center}
\end{figure}

In figure (\ref{fig:chi2 with silas}) we superimpose their limits on $B_{\D}$
upon our results from figure (\ref{fig:chi2 contours}).
Beane and Savage only considered pion masses smaller than its current value so while the two sets of curves seem to
overlap in a small area close to $\Delta \delta m_{np}/\delta m_{np} = \Delta B_{\D}/B_{\D} = 0$ a much larger overlap
might be expected if heavier pions were considered. The figure shows that even with the large systematic uncertainty in
deuterium BBN can rule out much of the region of increases in $\delta m_{np}$ and $B_{\D}$ up to $\Delta B_{\D}/B_{\D}
= 0.5$ because it is incompatible with the $B_{\D}$ - $\delta m_{np}$ relationship from Beane and Savage despite the
large uncertainty also found in that calculation. New limits will be found when heavier pions are considered, and also
when we have sufficient information to explore $\Delta B_{\D}/B_{\D} > 0.5$.


\section{Conclusions}
\label{sec:conclusions}

BBN presents a golden opportunity to study possible changes in the fundamental constants of nature, particularly those
related to the structure of nuclei and their interactions, at one of the earliest epochs of the Universe. We have
examined the impact of variations in one of these constants, $\Lambda_{QCD}$, upon the predictions for the primordial
deuterium abundance and helium-4 mass fraction. A change in $\Lambda_{QCD}$ will manifest itself through shifts in the
properties of the nuclei such as the neutron-proton mass difference and the deuteron binding energy. We have shown how
these two parameters are crucial for determining the predictions for \D~ and \He~ and how we can simplify BBN to the
extent that it is a function of only these two quantities. While the simplification gives very good predictions for the
mass fraction of \He~ when compared to a standard BBN code the results for \D~ were offset by 50\%. This large
systematic error in deuterium swamped the statistical error associated with the observation allowing the degeneracy
between $\delta m_{np}$ and $B_{\D}$ in the prediction of helium-4 to show through. By simulating the case when the
systematic error in the prediction for deuterium can be removed we found that BBN can limit their variation to the 10\%
level. In order to make BBN a better probe of the time variation of $\Lambda_{QCD}$, we need to know in particular, the
dependence of the $^3{\rm He}$ and ${\rm T}$ (and $^4{\rm He}$) binding energies on the pion mass and the dependence of
all the binding energies in the cross sections. Even without this input, much stronger constraints are obtained when
the result of the BBN calculation is combined with the results of the Beane and Savage calculations for the deuteron
binding energy. If $\Lambda_{QCD}$ is the only constant that varies with time, and we only consider increases in the
pion mass, then in order to be compatible with the results from Beane and Savage the variations of $B_{\D}$ and $\delta
m_{np}$ are limited to $\sim $ 1\%.


\vspace{1.cm} The authors would like to thank Rebecca Surman, Martin Savage
and Silas Beane for useful discussions. This work was supported by the
U.S. Department of Energy under grant DE-FG02-02ER41216.


\end{document}